\documentstyle[preprint,aps,pre]{revtex}
\tightenlines
\draft

\begin{document}

\title{Stochastic multiplicative processes with reset events}

\author{Susanna C. Manrubia}

\address{Fritz-Haber-Institut der Max-Planck-Gesellschaft\\ 
Faradayweg 4-6, 14195 Berlin, Germany}

\author{Dami\'an H. Zanette}

\address{Consejo  Nacional de  Investigaciones  Cient\'{\i}ficas  y
T\'ecnicas\\ Centro At\'omico Bariloche  e  Instituto  Balseiro\\
8400  S.C.  de Bariloche, R\'{\i}o Negro, Argentina}

\date{\today}

\maketitle

\begin{abstract}

We study a stochastic multiplicative process with reset events. It  is
shown  that  the  model  develops  a  stationary power-law probability
distribution for the relevant variable, whose exponent depends on  the
model parameters.  Two qualitatively  different regimes  are observed,
corresponding to intermittent and  regular behaviour. In the  boundary
between them, the mean value  of the relevant variable  is 
time-independent,
and  the  exponent  of  the  stationary  distribution equals $-2$. The
addition of diffusion to the system modifies in a non-trivial way  the
profile  of  the  stationary  distribution.  Numerical  and analytical
results are presented.

\end{abstract}

\pacs{PACS: 05.20.-y, 05.40.+j}

\newpage

The occurrence of power-law  distributions (PLDs) is a  common feature
in the description of natural phenomena. These distributions appear in
a  wide  class  of  nonequilibrium  systems,  ranging  from   physical
processes  such  as  dielectric  breakdown,  percolation,  and rupture
\cite{Gouyet}, to  biological processes  such as  dendritic growth and
large-scale evolution \cite{evol},  to sociological phenomena  such as
urban development \cite{urban}.  Power-laws have been associated  with
the effect of the complex driving mechanisms inherent to these systems
and with their intrincate dynamical structure.  Criticality, fractals,
and  chaotic  dynamics  are  known  to  be  intimately related to PLDs
\cite{fract}.

In view  of the  ubiquity of  PLDs in  the mathematical description of
Nature, much  work has  been recently  devoted to  detecting universal
mechanisms able to give  rise to such distributions.   In the frame  of
equilibrium processes,  for instance,  power-laws have  been shown  to
derive from  generalized maximum-entropy  formulations \cite{Tsallis}.
For  nonequilibrium  phenomena,  self-organized  criticality (SOC) and
stochastic  multiplicative  processes  (SMPs)  have been identified as
sources of  PLDs.   According to  the SOC  conjecture \cite{SOC}, some
nonequilibrium systems are continuously  driven by their own  internal
dynamics  to  a  critical  state  where,  as  for  equilibrium   phase
transitions,  power-laws  are  omnipresent.   On  the other hand, SMPs
\cite{SMP} provide  a (more  flexible) mechanism  for generating PLDs,
based in the presence of underlying replication events.

It  is  however  well  known  that  a  pure SMP,
\begin{equation}
n(t+1)=\mu(t) n(t),
\end{equation}
with $\mu$ a random variable,  does not generate a stationary  PLD for
$n(t)$.   Rather,  it  gives  rise  to  a  time-dependent   log-normal
distribution. To model the above mentioned phenomena, therefore,  SMPs
have to be combined with additional mechanisms. It has been shown that
transport  processes  \cite{tra},  sources  \cite{sou},  and  boundary
constraints \cite{bc} are able to induce a SMP to generate power-laws.
The aim of the present  paper is to discuss an  alternative additional
mechanism, namely,  randomly reseting  of the  relevant variable  to a
given  reference  value.  In  a  real  system,  this  would  represent
catastrophic  annihilation  or  death  events,  seemingly   originated
outside the system.

We consider a discrete-time stochastic multiplicative process  $n(t)$,
added with reset events in the  following way. At each time step,  $n$
is reset  with probability  $q$ to  a new  value $n_0$,  drawn from  a
probability  distribution  $P_0(n_0)$.  If  the  reset  event does not
occur,  $n$  is  multiplied  by  a  random  positive factor $\mu$ with
probability distribution $P(\mu)$. Namely,
\begin{equation}    \label{p0}
n(t+1) =  \left\{
\begin{array}{ll}
n_0(t+1) &\mbox{with probability $q$,} \\
\mu(t) n(t) &\mbox{with probability $1-q$}.
\end{array}
\right.
\end{equation}
Between two consecutive  reset events, $n(t)$  thus behaves as  a pure
multiplicative  process.   When  one   of  such  events  occurs,   the
multiplicative sequence starts again.

In order  to gain  insight into the dynamics  of process  (\ref{p0}) we
first  consider  the  simplest  case  where  $n_0(t)$ and $\mu(t)$ are
constant for all $t$. Since  an arbitrary factor in the  initial value
of $n$  is irrelevant  to its  subsequent evolution,  we take  $n_0=1$
without loss of generality. We have thus
\begin{equation}    \label{p1}
n(t+1) =  \left\{
\begin{array}{ll}
1 &\mbox{with probability $q$,} \\
\mu n(t) &\mbox{with probability $1-q$}.
\end{array}
\right.
\end{equation}
This stochastic recursive equation can be readily solved to give
\begin{equation}    \label{s1}
n(t)= \left\{
\begin{array}{ll}
\mu^k & \mbox{with probability $p_k=q(1-q)^k$ ($0\le k\le t-1$),}\\
\mu^t & \mbox{with probability $p_t=(1-q)^t$}.
\end{array}
\right.
\end{equation}
Note   that   the   possible   values   of   $n(t)$,   $n_k  =  \mu^k$
($k=0,1,...,t$), lie in  the interval $[\mu^t,1]$  for $\mu<1$ and  in
$[1,\mu^t]$ for $\mu>1$. Except for the extreme value $n_t=\mu^t$, the
associated probabilities  are time-independent.  As time  elapses, the
probability of each possible value of $n(t)$ is therefore quenched for
$n\neq \mu^t$, and the corresponding probability distribution  evolves
at this extreme value only. Thus, the distribution sequentially builds
up in zones that lie increasingly further from $n=1$.

For large times, when the number of possible values of $n(t)$  becomes
also large, it is possible to give the probability distribution $f(n)$
for $n\in (\mu^t,1]$ for $\mu<1$ and $n\in [1,\mu^t)$ for $\mu>1$ as a
function of a continuous variable. In fact,
\begin{equation}    \label{f1}
f(n)= p_k \left|\frac{dk}{dn}\right|= \frac{q}{|\ln \mu|}
n^{-\alpha},
\end{equation}
with  $\alpha  =  1-\ln(1-q)/\ln  \mu$.  In  order  to account for the
contribution at $n=\mu^t$,  $f(n)$ should be  added with a  delta-like
term $f_0(t) \delta(n-\mu^t)$, where the factor $f_0$ can be  obtained
from the normalization of $f(n)$.

According to (\ref{f1}), the stochastic process (\ref{p1}) gives  rise
to a {\it stationary power-law} distribution $f(n)$ in an increasingly
large interval  of values  of $n$.   For $t\to  \infty$, $f(n)$  is  a
stationary  power-law  distribution  in  $(0,1]$  for  $\mu<1$, and in
$[1,\infty)$ for $\mu>1$.   In contrast with multiplicative  processes
with boundary constraints  \cite{bc}, there are  no conditions on  the
parameters  to  obtain  a  stationary  power-law  distribution.    For
$1-q<\mu<1$,   the   exponent   of   this   distribution  is  positive
($\alpha<0$), and $f(n)$ grows with $n$.  In this situation,  however,
the distribution is defined for $0< n\le 1$ and exhibits a cut off  at
$n=1$.  On the  other hand, for $\mu<1-q$  or $\mu>1$ the exponent  is
negative ($\alpha>0$).  For $\mu>1$, i.e. when $n(t) \in  [1,\infty)$,
the  moments  $m_i  =  \int  f(n)  n^i  dn$  diverge for $i>\alpha-1$,
indicating    the    presence    of    intermittent     amplifications
\cite{tra,Mikh}. For $\mu<1-q$, $m_i$ diverges for $i<\alpha-1$.

It is interesting to relate the exponent of the power-law distribution
with the  evolution of  the mean  value $\langle  n(t)\rangle$.   From
(\ref{s1}), this mean value can be written as
\begin{equation}
\langle n(t)  \rangle = \frac{q}{1-(1-q)\mu}+
\frac{(1-q)(1-\mu)}{1-(1-q)\mu} \mu^t(1-q)^t.
\end{equation}
For $\mu(1-q)<1$, the mean value of $n(t)$ converges to a finite value
$\langle  n  \rangle  =q/[1-(1-q)\mu]$,  whereas  for  $\mu(1-q)>1$ it
``explodes''.   In the  boundary between  both regimes,  where $\mu  =
1/(1-q)$, the  exponent of  the distribution  is $\alpha=2$  and $f(n)
\sim n^{-2}$.  This exponent  is therefore  to be  associated with the
explosion threshold.

The power-law distribution in (\ref{f1})  can also be inferred from  a
description of the  evolution of $f(n)$.  In fact, since  at each time
step where no  reset occurs the  probability contribution to  $f(n)dn$
comes from $n'=n/\mu$, we can write
\begin{equation}    \label{e1}
f_{t+1}(n) dn= (1-q) f_t(n/\mu) d(n/\mu).
\end{equation}
Assuming now  that this  distribution is  stationary, $f_{t+1}  \equiv
f_t$, a solution to (\ref{e1}) is given by $f(n) =A n^{-\alpha}$, with
\begin{equation}
(1-q) \mu^{\alpha-1}=1,
\end{equation}
which produces the same value  of $\alpha$ as in Eq.  (\ref{f1}). Note
that (\ref{e1}) does  not hold for  $n=1$, where the  contributions to
the probability come from reset events.

The above  argument provides  a method  for dealing  with the  general
multiplicative process with reset  events, Eq.  (\ref{p0}),  when both
$\mu$ and  $n_0$ are  drawn from  prescribed probability distributions
$P(\mu)$  and  $P_0(n_0)$.  We  assume  that $P_0(n_0)$ is appreciably
different from zero in a bounded region, where the contributions  from
reset events are relevant. Outside this region the evolution of $f(n)$
can be written as
\begin{equation}    \label{e0}
f_{t+1} (n) dn = (1-q) \int_0^{\infty} d\mu\ P(\mu) f_t(n/\mu)
d(n/\mu),
\end{equation}
which   generalizes   Eq.   (\ref{e1}).   Under   the   assumption  of
stationarity, this equation is solved by $f(n) = An^{-\alpha}$,  where
the exponent $\alpha$ must verify
\begin{equation} \label{a}
(1-q) \int_0^{\infty} d\mu\ \mu^{\alpha-1}P(\mu) = 1.
\end{equation}
For regular forms of $P(\mu)$ this equation has at least one  solution
for $\alpha$. When the probability is mainly concentrated in values of
$\mu$  larger  than  unity  the  solution  is  expected to be positive
($\alpha>0$) and {\it vice versa}.

As in the case  of constant $\mu$ and  $n_0$, a close relation  exists
here between the  evolution of the  average $\langle n(t)\rangle$  and
the exponent  of the  power-law distribution.  In particular, $\langle
n(t)\rangle$ is  found to  remain stationary  along the  whole process
when $\alpha=2$.  Again, thus,  the exponent  $\alpha=2$ is associated
with the explosion threshold,  and marks the boundary  between regular
and intermittent evolution.  This can be seen, for instance, from  Eq.
(\ref{e0}). Multiplication  of this  equation by  $n$ and  integration
over $n$ yields
\begin{equation}    \label{en}
\langle n(t+1)\rangle = (1-q)\left[ \int_0^\infty d\mu\ \mu
P(\mu) \right] \langle n(t)\rangle.
\end{equation}
Comparing with Eq.  (\ref{a}), we readily note that the multiplicative
constant $(1-q)\int d\mu\  \mu P(\mu)$ that  governs the evolution  of
$\langle n(t)\rangle$ in Eq. (\ref{en}) equals unity for $\alpha=2$.

In summary,  depending on  $q$ and  $P(\mu )$  the system  can be in a
regular  regime  where  $\langle  n  (t)\rangle$ converges to a finite
value, or  in an  intermittence regime,  where $\langle  n(t) \rangle$
diverges. At the boundary,  i.e. at the explosion  threshold, $\langle
n(t)  \rangle$  remains  constant  and,  independently of the specific
value of $q$ and of the  particular form of $P(\mu )$, the  probability
distribution $f(n)$  exhibits a  power-law tail  with a characteristic
exponent, $f(n) \sim n^{-2}$.

We  have  numerically  checked  that  the  exponent  of the stationary
profile  of  $f(n)$  does  not  depend  on  the particular form of the
distribution of  reset values  $P_0(n_0)$. In  Fig. 1  we present  the
function  $f(n)$  obtained  with  constant  $\mu$  and  $q$, for three
different  choices  of  $P_0(n_0)$:   A  uniform  distribution between
$n_0=0$   and   $n_0=1$   (circles),   an   exponential  distribution,
$P_0(n_0)=\langle n_0\rangle^{-1}\exp (-n_0/\langle n_0\rangle)$, with
$\langle  n_0  \rangle=100$  (squares),  and  a  discrete distribution
$P_0(n_0)= [\delta(n_0-1)+\delta(n_0-100)]/2$.  The particular form of
$P_0$ sets a lower boundary for  the region where $f(n)$ behaves as  a
power law, but does not  affect the corresponding exponent.  Solid
lines  in  the  log-log  plot  of  Fig.   1 have the theoretical slope
$\alpha=1.1054...$. 

Figure 2  shows our  simulation results  for three  different forms of
$P(\mu)$:  An exponential distribution $P(\mu)=\langle\mu \rangle^{-1}
\exp(-\mu/\langle  \mu  \rangle  )$  with  $\langle  \mu\rangle=2$,  a
uniform distribution $P(\mu)=5/2$ with  $\mu \in [9/10,13/10]$, and  a
discrete distribution $P(\mu)= \sum_{k=1}^3 \delta(\mu-\mu_k)/3$  with
$\mu_1=1$, $\mu_2=6/5$ and $\mu_3=7/5$.  The slope of the solid  lines
has  been  obtained  numerically  for  various  values of $q$ from Eq.
(\ref{a}).   This  yields   $\alpha=1.4965...$  for  the   exponential
distribution  with   $q=0.2$,  $\alpha=1.2195...$   for  the   uniform
distribution with  $q=0.02$, and  $\alpha=1.8965...$ for  the discrete
distribution  with  $q=0.15$.   In   all  cases,  our  numerical   and
analytical results are in full agreement within six to nine decades in
the power-law region.

We have also  investigated the effects  of diffusive transport  on the
process   (\ref{p1}).   With   this   aim,   we   have   considered  a
one-dimensional array of elements  whose individual dynamics is  given
by  (\ref{p1})  and,  at  each  time  step,  we  have  incorporated an
interaction mechanism that mimics diffusion.  After the multiplicative
process with reset events has been applied, the state of each  element
is further changed to
\begin{equation}    \label{dif}
n'_i (t) = (1-D) n_i(t)
+ {D \over 2} \left[ n_{i+1}(t)+n_{i-1}(t)\right] ,
\end{equation}
where $i$  labels the  elements in  the array,  with periodic boundary
conditions. Then, $n'_i$ is used as the input state for the next step.
In this deterministic,  time-discrete version of  diffusive transport,
$D$ plays the role of a diffusion constant.

Figure 3 summarizes our numerical  results on the effect of  diffusion
on  the  SMP  (\ref{p1}),  displaying  the dependence of the power-law
exponent with the diffusion constant. We have chosen values of $q$ and
$\mu$  such  that  the  different  regimes  of  the  process have been
explored. The value of the  multiplicative constant has been fixed  in
this  case  to  $\mu=4/3$.  In  the  regular  regime  (i.e.
$\mu(1-q)<1$),  diffusion  produces  a  decrease  of  $\alpha$  in the
power-law distribution. This can be understood if we consider that the
role of diffusion is to deplete dense areas, transporting material  to
less occupied cells. The multiplicative process is not fast enough  in
this regime to balance the joint effect of reset events and diffusion.
As a result,  underpopulation occurs in  the high-density region,  and
$\alpha$ decreases ($q=0.3$  in Fig. 3).   In the intermittent  regime
($q=0.23$ i.e.  $\mu(1-q)>1$), diffusion  favors the  opposite effect.
Remarkably,  diffusion  does  not  have  any  effect  on  the value of
$\alpha$  when  the  system  is  evolving  at the explosion threshold.
Within numerical  errors, in  fact, $\alpha=2$  irrespectively of  the
value of  $D$. It  is also  worth to  point out  that the  qualitative
behaviour of the process depends  on $\mu$ and $q$ only.  Changing $D$
does not allow the system  to switch between the intermittent  and the
regular regimes.

Summing up, in this paper we have studied a stochastic  multiplicative
process with  reset events.  The combination  of this  random reseting
with the replication  events driven by  the stochastic process  allows
for the development of a  stationary distribution in the system,  both
when the  mean value  of the  relevant variable  converges to  a finite
value (regular  regime) and  when it  diverges (intermittent regime).
The regime at the boundary between regular and intermittent  behaviour
is of particular interest. At this point, where the overall effects of
the multiplicative process are exactly balanced by the random  resets,
the mean value of the  relevant variable remains constant in  time. We
have shown that  this property is  closely related with  the fact that
the exponent  of the  power-law stationary  distribution equals  $-2$.
This value is  to be related  with Zipf law,  which predicts the  same
exponent of power-law distributions in a series of seemingly disparate
natural systems  \cite{evol,urban}. Thus,  the SMP  with reset  events
offers  an  alternative  explanation  of  this ubiquitous exponent. In
fact, whereas a general trend  of biological and social systems  could
be to improve their growth rates by increasing the parameter $\mu$, it
is on the other hand to be expected that external constrains are going
to operate in order to avoid divergencies by increasing $q$. It is not
unlikely that the competition  between these two processes  could lead
real systems to this  boundary between regular behavior  and developed
intermittency.

\section*{Acknowledgements}

Financial support from Fundaci\'on Antorchas, Argentina, and from 
Alexander von Humboldt Foundation, Germany (SCM) is gratefully
acknowledged. 



\begin{figure}
\caption{Stationary distribution  $f(n)$ for  $\mu=1.1$ and  $q=0.01$,
and for different distributions of reset values $P_0(n_0)$ (see text).  
Straight  lines have the theoretical  slope $\alpha=1.2195\dots$.} 
\end{figure}

\begin{figure}
\caption{Stationary  distributions  $f(n)$  for  different  forms   of
$P(\mu)$ and  different values  of $q$  (see text).  The slope  of the
straight  lines  has  been  obtained  through  numerical  solution  of
(\ref{a}).}

\end{figure}

\begin{figure}
\caption{Dependence  of  the  exponent   $\alpha$  on  the   diffusion
coefficient $D$ for $\mu=4/3$ and three values of $q$ corresponding to
the intermittence regime ($q=0.23$), the regular phase ($q=0.3$),  and
the explosion threshold ($q=0.25$). The error bars stand for the error
of $\alpha$ in a least square fit to the numerical data.}

\end{figure}

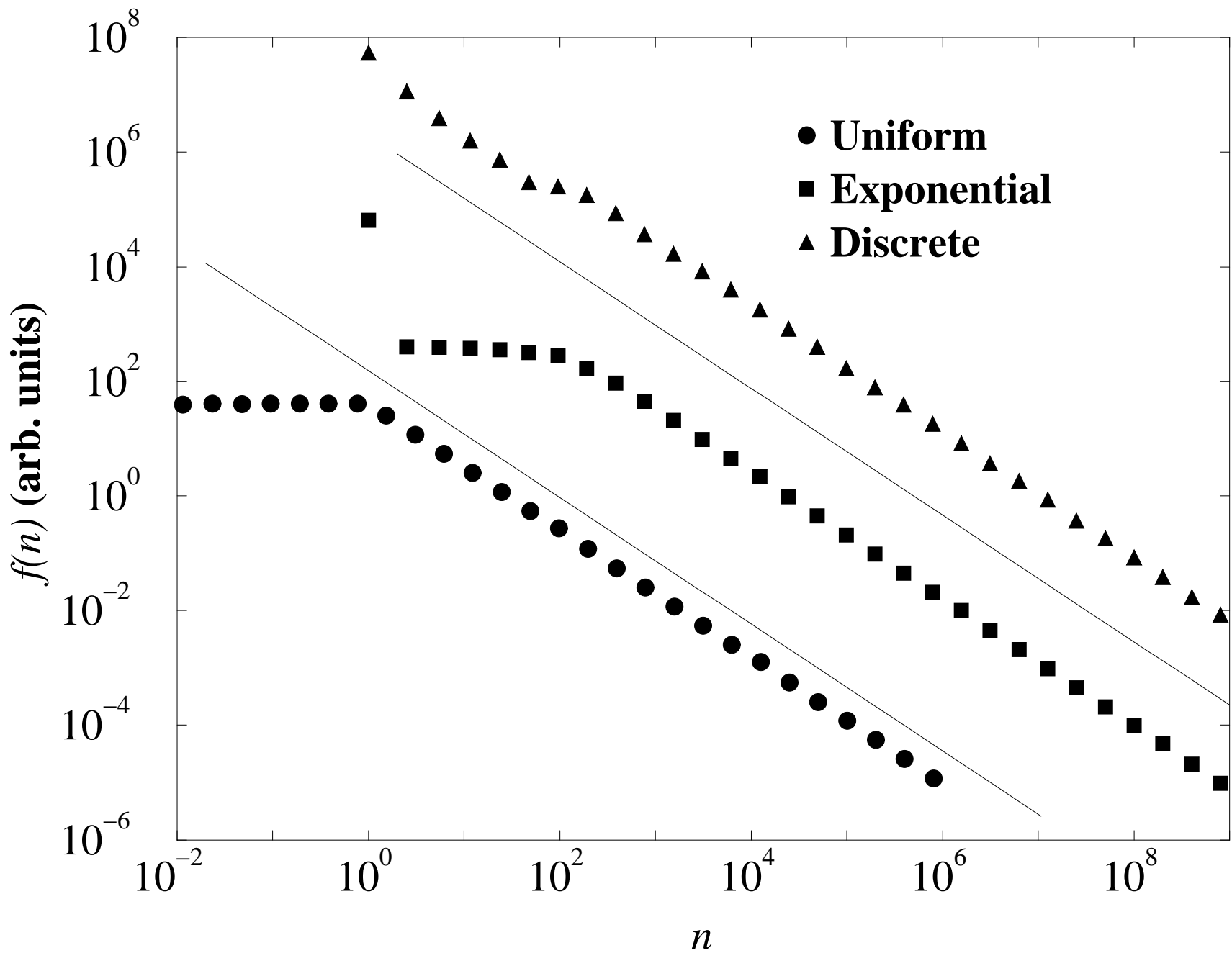
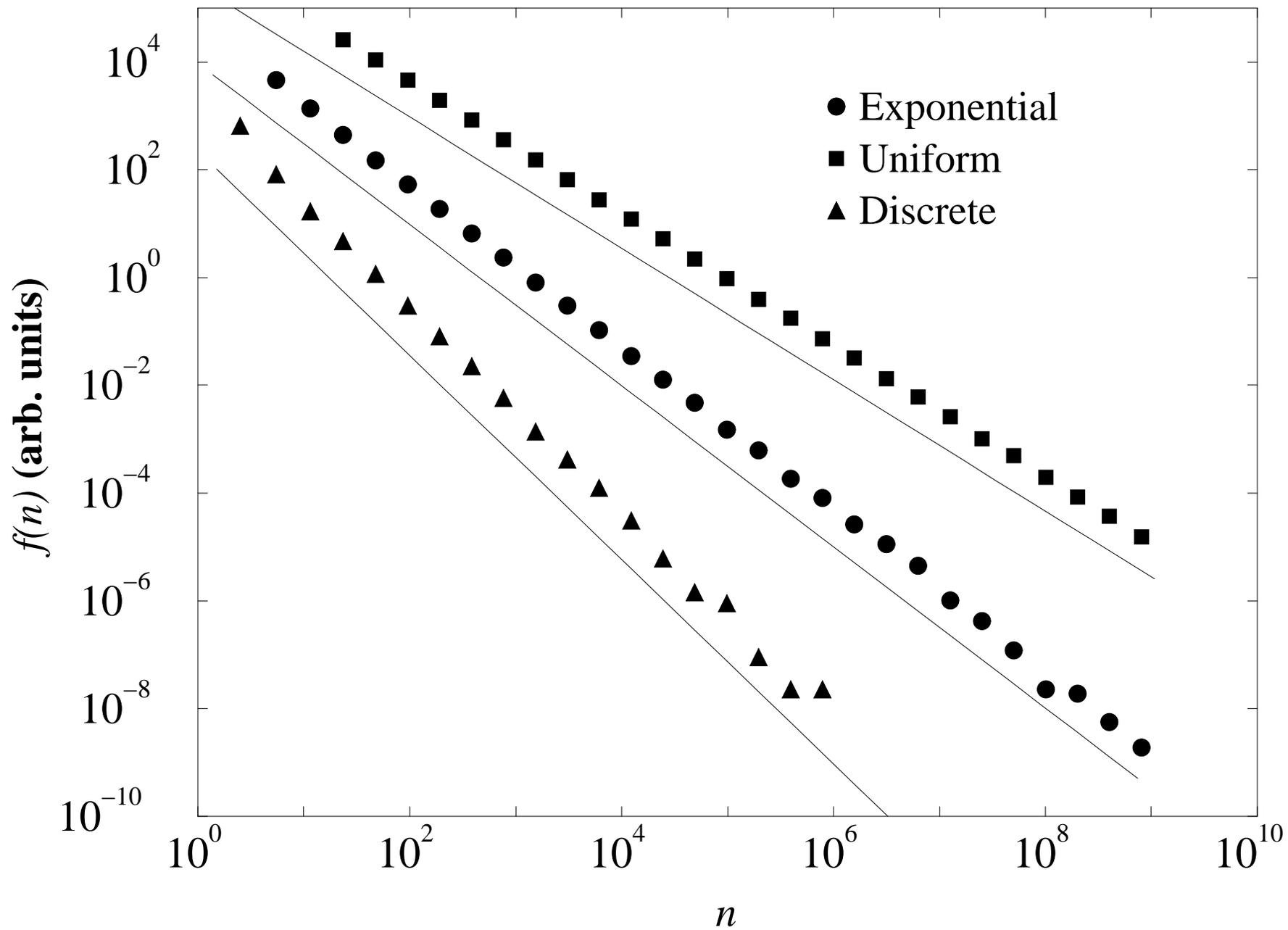
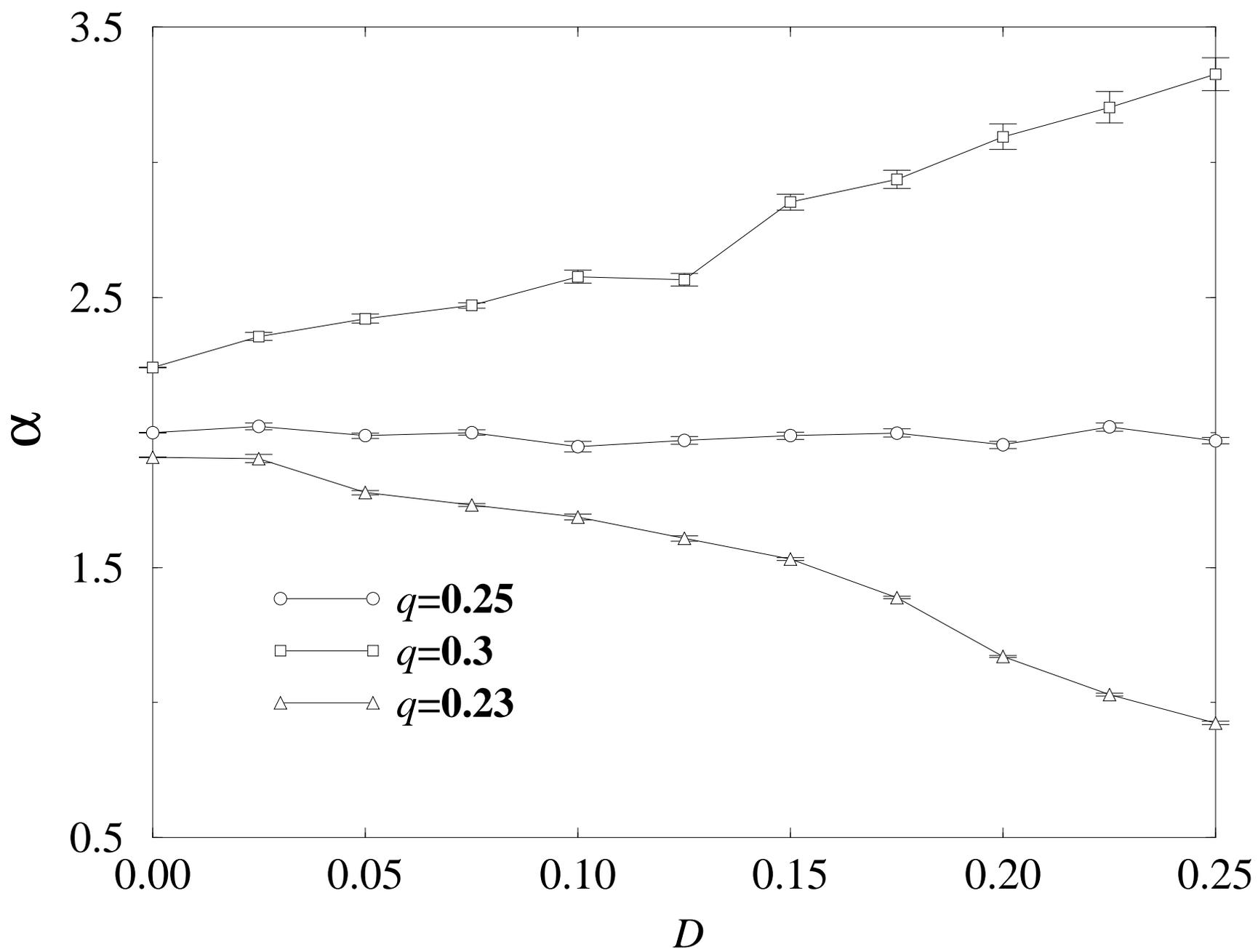
\end{document}